\documentclass[
  reprint,
  superscriptaddress,
  amsmath,
  amssymb,
  aps,
  prb
]{revtex4-2}

\usepackage{graphicx}
\usepackage{dcolumn}
\usepackage{bm}
\usepackage{hyperref}

\begin{document}

\title{Comment on ``Topography of Fermi arcs in \textit{t}-PtBi$_2$ using high-resolution angle-resolved photoemission spectroscopy''}

\author{Sergey Borisenko}
\email{S.Borisenko@ifw-dresden.de}
\affiliation{Leibniz IFW Dresden, Helmholtzstrasse 20, 01069 Dresden, Germany}

\date{\today}

\begin{abstract}
Angle-resolved photoemission spectroscopy (ARPES) and scanning tunneling spectroscopy have established surface superconductivity in the Weyl semimetal \textit{t}-PtBi$_2$, whereas O'Leary \textit{et al.} concluded from an independent ARPES experiment that superconducting signatures are absent above 3 K. Here we reanalyze the complete deposited data of O'Leary \textit{et al.} Using only experimentally determined momentum-distribution-curve (MDC) maxima, energy-distribution-curve (EDC) peak positions, and leading edges, we find a strongly anisotropic low-energy gap and dispersion back-bending that closely reproduce the magnitude and angular dependence reported independently by Changdar \textit{et al.} We further show that the temperature-dependent data of Ref.~\cite{Oleary} originate from the other surface termination and were presented using temperature-dependent energy translations, momentum translations, and momentum rescalings. Registering the original temperature-dependent datasets using the metallic bulk Fermi cutoff reveals finite gaps of approximately 2, 2, 1, and 0.8 meV at the four Fermi crossings. Thus the independently acquired data of O'Leary \textit{et al.}, obtained on different samples and with a different photon energy, confirm rather than contradict the previously reported anisotropic superconducting gap and additionally show that an anisotropic low-energy suppression persists to 19 K.
\end{abstract}

\maketitle

Fermi arcs are metallic surface states characteristic of Weyl semimetals. At the Fermi level, they form open contours in momentum space terminating at the projections of bulk Weyl points. On approaching a projected Weyl point, a surface state can lose its distinct surface character as it merges with the projected bulk continuum. Such merging, however, does not by itself open an energy gap at a Fermi crossing. Indeed, the calculated surface-state dispersions presented by O'Leary \textit{et al.} in Figs.~1(e)--1(g) and 2(e) cross \(E_F\) at the corresponding momenta \cite{Oleary}.

A superconducting gap produces qualitatively different signatures in ARPES. The spectral peak and leading edge are displaced to higher binding energies at and near the normal-state Fermi momentum, while the occupied dispersion develops the characteristic back-bending associated with particle-hole mixing. For an anisotropic order parameter, these effects vary systematically along the Fermi surface and become minimal at nodes. Consequently, spectral peak positions, leading edges, and dispersion topology provide more direct diagnostics of a gap than absolute photoemission intensity, which can vary strongly through matrix-element effects.

O'Leary \textit{et al.} reported detailed ARPES measurements of the Fermi arcs of \textit{t}-PtBi$_2$ and concluded that their low-temperature data show neither a superconducting gap nor the corresponding Bogoliubov dispersion \cite{Oleary}. Since the underlying datasets are publicly available \cite{DataLink}, these conclusions
can be tested directly without modeling the spectral function. Here we
do so by tracking only the positions of experimentally observed spectral
features. EDC and MDC peak positions are obtained from local fits to
their maxima, while leading-edge positions are determined from local
fits to the maximum of the EDC derivative, $dI/dE$. Simple Gaussian
or Lorentzian functions are sufficient for these local fits and yield
indistinguishable peak positions within the experimental precision.
 Only a restricted region around the corresponding feature was used, so that these fits serve to determine its position rather than to model the full spectral lineshape. Importantly, leading-edge positions are also the principal spectral quantity used by O'Leary \textit{et al.} themselves to distinguish crossing from non-crossing branches of the Fermi arcs; our analysis therefore extends the procedure already applied in Ref.~\cite{Oleary} to the remaining cuts of the deposited dataset.

\section{Gap anisotropy}

Figure~1(a) reproduces the relevant portion of the Fermi-surface map from Fig.~4(a) of Ref.~\cite{Oleary}. The black lines indicate the positions of the eight cuts shown in Figs.~4(c)--4(j) of that work. We additionally include the cut presented in Fig.~3(b). All cuts intersect the Fermi arc at well-defined positions.

Our analysis closely follows the procedure adopted by O'Leary \textit{et al.} to establish the Fermi-arc character of these states. In their Fig.~3, the leading-edge positions of EDCs at the two sides of the surface band are used to determine whether the corresponding branches reach $E_F$. In particular, the leading edges of the two crossing EDCs in Fig.~3(c) differ by approximately 1.52~meV, and this difference forms part of the evidence that one side does not cross $E_F$, establishing the open character of the Fermi contour. Here we extend the same leading-edge analysis to the additional cuts through the same Fermi-surface dataset shown in Fig.~4.

Because the relevant energy differences are on the meV scale, we independently evaluate the relative energy reproducibility directly from the deposited data. Figure~S1 compares the leading edge of the common bulk-derived metallic spectral weight in all eight independently acquired cuts underlying Fig.~4 of Ref.~\cite{Oleary}. Over detector channels 79--234, a broad interval selected from the raw spectra while excluding the Fermi-arc features, the rms deviation of the individual leading-edge positions from their point-by-point mean trajectory is 0.272~meV. Of the 1248 individual residuals, 95\% satisfy $|\delta E|<0.557$~meV, while the mean offsets of the individual cuts do not exceed 0.154~meV in magnitude. The latter directly constrains acquisition-to-acquisition relative energy shifts. These empirical variations are substantially smaller than both the 1.52-meV leading-edge difference used in Ref.~\cite{Oleary} and the several-meV angular variation discussed below.

We label the crossings of the eight cuts by numbers and denote the central crossing of the Fig.~3(b) cut by N, corresponding to the nodal direction in the notation of Ref.~\cite{Changdar}. The second crossing of this cut, where the surface state approaches the projected bulk states, is denoted B. The crossing positions are parameterized by an angle $\phi$ around the arc, with $\phi=0$ at N and positive angles defined clockwise, following the convention of Ref.~\cite{Changdar}.

We determine the characteristic energy at every crossing using three independent procedures. First, for each cut we determine the positions of the MDC maxima at $E_F$ from local fits and extract EDCs at the corresponding momenta. The leading-edge position can equivalently be estimated directly from the midpoint of the EDC leading edge; fitting the maximum of \(dI/dE\) provides a more precise and noise-robust determination. The leading-edge positions of these EDCs provide the first estimate of the gap-related energy scale. All 18 EDCs used in this analysis are displayed in the left panel of Fig.~1(c). Their systematic displacement along the arc is already apparent in the raw spectra.

Second, we determine the leading-edge position for every EDC from a local fit. The resulting leading-edge energy as a function of momentum is shown in the middle panel of Fig.~1(c). Local extrema near the expected Fermi crossings provide a determination that does not depend on selecting a particular EDC from the finite width of an $E_F$-MDC maximum.

Third, we determine the EDC peak energy throughout each cut from local fits to the spectral maximum. The corresponding peak-position curves are shown in the right panel of Fig.~1(c). For most cuts the extrema associated with both crossings are directly resolved. For cuts $i$ and $j$, the increasing bulk spectral contribution between the crossings shifts and broadens these extrema. At these positions we therefore use the EDCs identified by the leading-edge analysis and assign correspondingly larger uncertainties.
The uncertainties in Fig.~1(b) include the empirical relative-energy reproducibility determined from Fig.~S1 in addition to the uncertainty associated with locating the corresponding spectral feature.

The three procedures yield the angular dependences shown in Fig.~1(b). Despite their different sensitivities to linewidth, momentum resolution, and particle-hole mixing near $k_F$, all three show the same qualitative behavior: the energy scale is minimal near N, increases strongly away from this direction, reaches approximately 3--5 meV depending on the estimator, and decreases again on approaching the projected Weyl-point regions.

For comparison, Fig.~1(b) also shows the gap anisotropy reported independently in Ref.~\cite{Changdar}. The agreement is remarkable. The data of O'Leary \textit{et al.} reproduce both the nodal behavior and the characteristic magnitude of the previously reported anisotropic gap, with only a modest angular displacement of the maxima.

This agreement is particularly significant because the measurements were performed independently, on different samples and ARPES systems, and using substantially different low photon energies. Ref.~\cite{Changdar} employed approximately 6-eV excitation, whereas O'Leary \textit{et al.} used approximately 7-eV photons \cite{Oleary}. Such a change generally modifies photoemission matrix elements and relative spectral intensities considerably. Nevertheless, the independently acquired datasets yield essentially the same momentum dependence and magnitude of the low-energy gap. Moreover, the present evidence does not rely on absolute intensity: the gap is independently encoded in leading-edge positions, EDC peak positions, and, as shown below, dispersion back-bending.

\section{Back-bending of the dispersion}

A second conclusion of Ref.~\cite{Oleary} is that the measured dispersions do not exhibit the back-bending expected from superconducting Bogoliubov quasiparticles. The deposited data allow this statement to be tested directly.

Figure~2(a) shows five representative momentum cuts, including data presented in Figs.~4(c), 4(e), and 4(f), as well as additional deposited cuts associated with the same measurements. For every EDC between the two Fermi crossings we determine the peak and leading-edge positions from the same local fitting procedure and superimpose the resulting trajectories on the raw intensity maps.

Both quantities exhibit a reversal of dispersion near the Fermi crossings. The magnitude of the back-bending is typically of order 1--2 meV away from the nodal direction, whereas it becomes very small for the nodal cut of Fig.~3(b) of Ref.~\cite{Oleary}, consistent with an anisotropic gap that becomes minimal or zero at this position.

Figure~2(b) displays the corresponding raw EDC sequences for one crossing from each of the cuts in Fig.~2(a). The reversal is therefore not a consequence of image contrast, interpolation, or a particular color scale; it follows directly from the evolution of the spectral maxima in the deposited EDCs.

The simultaneous occurrence of a leading-edge displacement and dispersion back-bending near all crossings provides a second, independent signature of gap formation in the data of Ref.~\cite{Oleary}. Merging of a Fermi-arc state with projected bulk states may broaden the surface feature but it does not account for a systematic reversal of dispersion below $E_F$ at the experimentally determined Fermi crossings.

\section{Temperature-dependent measurements}

The temperature-dependent measurements presented in Fig.~5 of Ref.~\cite{Oleary} form a central part of the argument for the absence of superconductivity above 3 K. Reanalysis of the complete deposited datasets reveals three issues that are essential for interpreting this comparison: the surface termination, the momentum registration, and the energy registration.

\subsection{Surface termination}

The temperature-dependent measurements in Fig.~5 are not taken from termination A investigated in the preceding figures. This follows directly from the momentum-space dimensions and geometry of the arcs. The arc in Fig.~5 is at least a factor of two larger than any possible cross section of the termination-A arcs in the maps and cuts presented earlier in Ref.~\cite{Oleary}. Moreover, none of the cuts from Fig. 4 look qualitatively similar.

Figure~3(a) quantifies this distinction using the arc width $A$ and the separation $B$ between neighboring arcs. For termination A the ratio $A/B$ remains below approximately 0.15, whereas for the Fig.~5 datasets we obtain 0.18 at 3 K and 0.21 at both 11 and 19 K. Comparison with the calculated Fermi-surface maps for the two terminations from Ref.~\cite{Veyrat}, shown in Fig.~3(b), unambiguously identifies the Fig.~5 measurements with the other surface termination. Thus the temperature series used in Ref.~\cite{Oleary} to assess the superconducting gap does not probe the termination characterized in the preceding gap analysis.

\subsection{Momentum registration}

All three deposited temperature-dependent datasets have identical dimensions of 246 energy by 380 angular pixels \cite{DataLink}. Nevertheless, the momentum calibration used for the panels of Fig.~5 is temperature dependent. For 3, 11, and 19 K, respectively, the published momentum scalings are
\begin{align}
k_{\rm start} &= -0.0398931,\ -0.0301382,\ -0.0273353,\\
\Delta k &= 0.00137562,\ 0.00148768,\ 0.00153122.
\end{align}
Thus, in addition to independent translations of the momentum origins, the momentum increment per pixel increases by approximately 8.1\% at 11 K and 11.3\% at 19 K relative to the 3 K dataset. We find no description of these temperature-dependent momentum rescalings in Ref.~\cite{Oleary}.

However, the complete datasets do not become globally registered even after these transformations. As seen in Fig.~3(a), the arc width \(A\) is significantly smaller at 3 K than at 11 and 19 K. The mismatch is even more apparent for the crossings of the neighboring arc centered near $0.45~\text{\AA}^{-1}$, marked by the vertical red dashed lines in Fig.~3(a). The 3 K dataset also exhibits a distinctly different intensity distribution near the bottoms of the arcs.

These differences can be examined independently of any momentum calibration by using the original angular-pixel coordinates. Figure~3(c) shows the EDC peak-position (PP) and leading-edge (LE) trajectories extracted from the three deposited datasets in their native pixel coordinates. Both the shapes and relative positions of the trajectories show that the 3 K measurement follows a different momentum path. Neither the PP nor the LE trajectories can be brought into even approximate global coincidence by a simple translation. Notably, in the raw coordinates it is the right crossing, near $0.1~\text{\AA}^{-1}$ in Fig.~3(a), that remains relatively stable with temperature, whereas the momentum transformation used in Ref.~\cite{Oleary} aligns the left crossing and additionally rescales the momentum axis.
The EDCs used in Ref.~\cite{Oleary} as the Fermi-crossing spectra occur at angular pixels 29, 19, and 18 for 3, 11, and 19 K, respectively, whereas the EDCs used as the presumed band-bottom references occur at pixels 67, 63, and 57. Their separations are therefore 38, 44, and 39 pixels, respectively. Thus, the two sets of selected spectra do not maintain a common separation even in the original detector coordinates, independently of the subsequent momentum rescaling.
Direct determination of the minima from the complete EDC peak-position trajectories shows that the 19 K reference EDC is located at the band minimum, whereas the spectra selected at 3 and 11 K are not. Consequently, the spectra used as internal references do not represent equivalent momentum positions within the three measured dispersions.
The 3 K dataset additionally contains a replica of the left arc that is absent at higher temperatures, consistent with a change in the illuminated sample region. Together with the changes in arc geometry and relative crossing positions, these observations show that the temperature series cannot be treated as repeated measurements along an identical momentum trajectory without explicit registration. A dynamic comparison of the near-\(E_F\), full-angular-range deposited arrays in their native coordinates is provided in the Supplemental Material.
\subsection{Energy registration}

The energy calibration also changes between the three datasets. In Figs.~5(a)--5(c), the common energy increment is 0.89097 meV per pixel, but the starting energies are
\begin{equation}
E_{\rm start}=-155.4,\ -154.7,\ -154.0\ {\rm meV}
\end{equation}
for 3, 11, and 19 K, respectively. Relative to the 19 K dataset, these correspond to energy translations of 1.4 and 0.7 meV for 3 and 11 K. In subsequent EDC panels of Fig.~5, additional temperature-dependent offsets are used; for example, the 3 and 11 K EDCs have starting energies of $-155.8$ and $-154.5$ meV, respectively. Thus a single common energy transformation is not used throughout the different panels of Fig.~5.

Ref.~\cite{Oleary} recognizes the need for energy alignment and uses an EDC near the minimum of the shallow surface-state band as an internal reference. As shown above, however, the selected EDCs do not consistently correspond to the actual minima of the respective dispersions. Moreover, there is no obvious physical reason to regard a shallow surface-state feature located only about 6 meV below $E_F$ as an invariant energy reference between 3 and 19 K. We therefore determine the relative energy registration independently.

Between the surface arcs the deposited datasets contain broad bulk-derived metallic spectral weight with a well-defined Fermi cutoff [Fig.~3(a)]. We integrate a 155-pixel-wide region between the left and right arcs and use this cutoff solely to determine the relative energy displacement between the 3, 11, and 19 K measurements. This procedure makes no assumption concerning the temperature dependence of the shallow surface-state band. The absolute energy reference is retained from O'Leary \textit{et al.}: at 19 K, $E_F$ intersects the leading edge of the $k_F$-EDC at approximately 75\% of its maximum intensity. The latter correspond to $E_F$-MDC maxima for all crossings. Thus, our analysis does not introduce a new absolute energy reference. The 3 and 11 K datasets are then registered relative to 19 K using the bulk Fermi cutoff, without any further adjustment of individual EDCs.

Figure~3(d) compares the spectra after the relative registration determined from the bulk Fermi cutoff and the single absolute surface-state reference described above. All EDCs are labeled by their original pixel numbers so that the extraction can be reproduced directly from the deposited arrays. We perform the analysis on the deposited raw datasets; in particular, no smoothing is applied to the EDCs shown in Fig.~3. The EDCs displayed in Figs.~4 and ~5 of Ref.~\cite{Oleary} appear to have been smoothed, whereas all EDCs shown here are extracted directly from the deposited raw arrays without smoothing.

The resulting temperature evolution is systematic at all four Fermi crossings. The correctly registered spectra imply low-temperature gap values of approximately 2 meV, 2 meV, 1 meV, and 0.8 meV for the four crossings, respectively. Upon cooling, the leading edges move away from the surface-state Fermi reference and low-energy spectral weight is suppressed. These changes are obtained without any further adjustment of individual spectra.

An additional observation follows directly from the same analysis. Since the absolute energy reference is fixed by assigning the near-nodal 19 K crossing to a gapless Fermi cutoff, the remaining crossings provide an independent test of the high-temperature state. The other crossings retain finite leading-edge shifts already at 19 K. Thus the anisotropic low-energy suppression does not vanish at the highest temperature measured by O'Leary \textit{et al.}; rather, its momentum dependence persists to 19 K and evolves continuously on cooling. This behavior is consistent with the pseudogap-like state observed in our independent measurements \cite{Changdar}, although establishing its microscopic origin is beyond the scope of the present Comment.

\section{Discussion and conclusion}

Our reanalysis uses only the publicly deposited measurements of O'Leary \textit{et al.} and relies on spectral quantities obtained by local fits directly to the deposited data \cite{DataLink}: MDC maxima, EDC peak positions, and leading-edge positions. No superconducting spectral-function model is required.

The analysis furthermore does not introduce a new criterion for identifying whether a surface-state branch reaches $E_F$. It extends the leading-edge procedure used in Ref.~\cite{Oleary} itself to establish the open Fermi-arc contour. An internal comparison of the eight independently acquired cuts gives an rms relative-energy reproducibility of 0.272~meV and mean cut-to-cut offsets below 0.154~meV, substantially smaller than the energy scales discussed here.

Three independent observations emerge. First, the low-temperature data exhibit a strongly anisotropic gap whose magnitude and angular dependence closely reproduce those reported independently in Ref.~\cite{Changdar}. Second, the same datasets show dispersion back-bending at the corresponding Fermi crossings, with the effect becoming minimal in the nodal direction. Third, after relative registration of the temperature-dependent datasets using their bulk metallic Fermi cutoff, the spectra show a systematic low-temperature gap at all four Fermi crossings, with values of approximately 2, 2, 1, and 0.8 meV.

The agreement with the earlier measurements is particularly notable because it is obtained from different samples, a different ARPES setup, and a substantially different photon energy. A change from approximately 6 to 7 eV strongly modifies low-energy photoemission matrix elements and relative intensities. Nevertheless, the independently acquired data reproduce the same anisotropic low-energy scale. Furthermore, the agreement is found in peak and leading-edge energies and in dispersion back-bending rather than merely in absolute spectral weight. It therefore cannot be naturally attributed to a common photoemission matrix-element effect.

Conversely, the apparent temperature independence emphasized in Fig.~5 of Ref.~\cite{Oleary} is based on measurements from the other surface termination and on datasets presented with temperature-dependent momentum translations, momentum rescalings, and energy offsets. The surface-state EDCs used as internal energy references are not consistently located at the corresponding band minima. The published momentum increments differ by 8.1\% and 11.3\% from that of the 3 K dataset at 11 and 19 K, respectively, despite identical array dimensions. These transformations are consequential because the temperature-dependent dispersions do not coincide globally: the selected curves are brought into agreement only near the particular Fermi-crossing spectrum used for the principal comparison, while remaining displaced elsewhere.

When the original datasets are instead registered relatively using the bulk metallic Fermi cutoff, and a single absolute surface-state energy reference is fixed from the near-nodal 19 K spectrum, all remaining spectral shifts become outcomes rather than alignment conditions. The off-nodal crossings remain gapped already at 19 K, and their gap-related energy scales increase upon cooling. The persistence of the anisotropic suppression to 19 K suggests a pseudogap-like regime above the temperature of coherent superconductivity and provides an independent reproduction of the corresponding behavior observed previously.

We therefore conclude that the deposited data of O'Leary \textit{et al.} \cite{DataLink} provide independent evidence for superconductivity of the Fermi arcs in \textit{t}-PtBi$_2$, rather than evidence for its absence above 3 K. More generally, the reanalysis illustrates that temperature-dependent ARPES comparisons at the meV scale require a common and independently verifiable registration of both momentum and energy coordinates, particularly when the claimed effect is comparable to the applied coordinate corrections.

\section*{Acknowledgments}
We are grateful to Susmita Changdar, Andrii Kuibarov, Jeroen van den Brink, and Bernd Büchner for fruitful discussions.

\begin{figure*}
    \centering
    \includegraphics[width=1\linewidth]{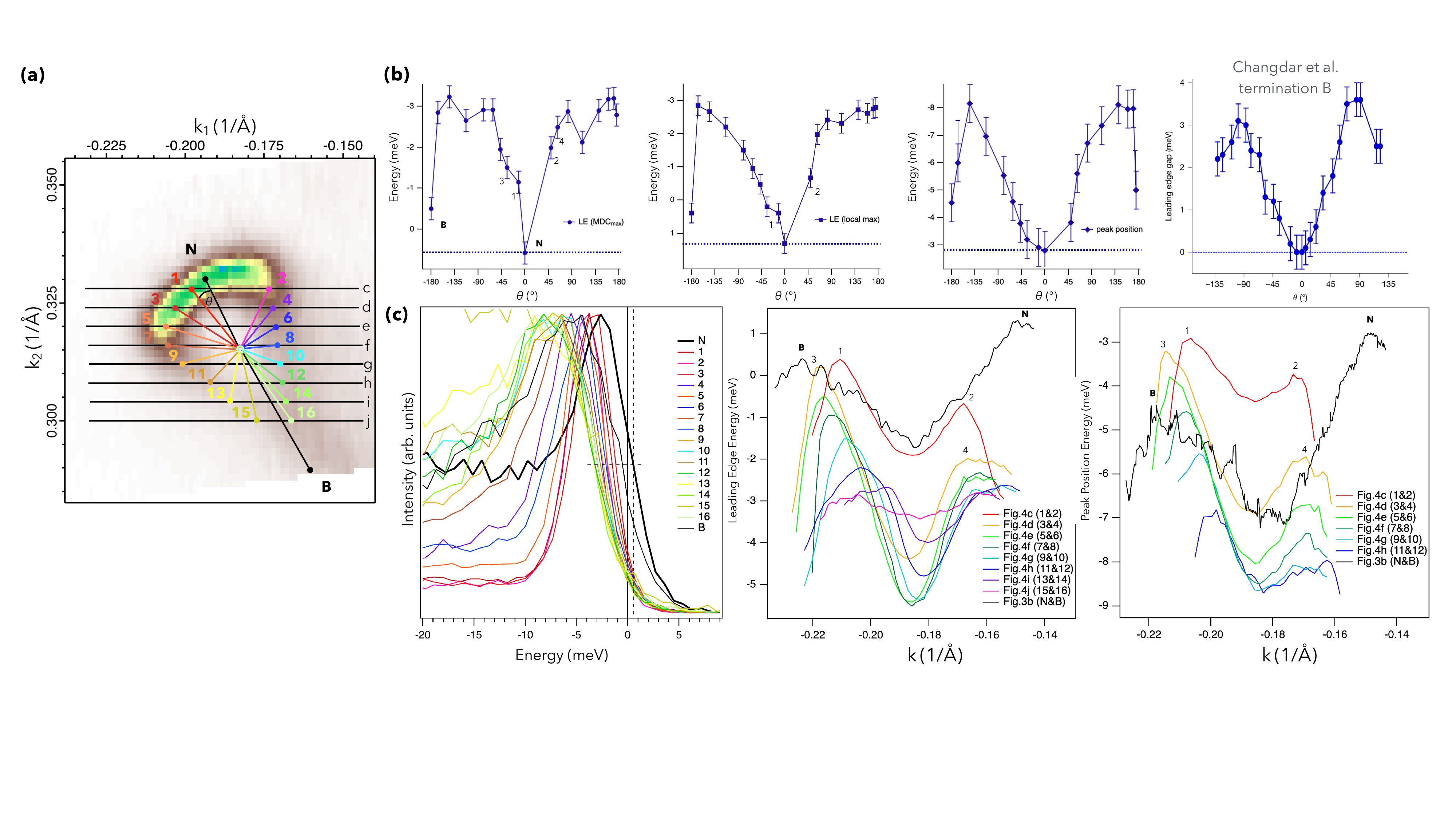}
    \caption{
\textbf{Anisotropic gap in the data of O'Leary \textit{et al.}}
(a) Portion of the Fermi-surface map of termination A from
Ref.~\cite{Oleary}. Black lines indicate the momentum cuts shown in
Figs.~4(c)--4(j) of Ref.~\cite{Oleary}; the additional cut through N
and B is taken from their Fig.~3(b). The intersections with the Fermi
arc are numbered 1--16. N denotes the crossing closest to the nodal
direction in the notation of Ref.~\cite{Changdar}, and B the second
crossing of the same cut. The angle $\theta$ parameterizes position
along the arc, with $\theta=0$ at N.
(b) Angular dependence of the low-energy scale extracted from the
deposited data using three procedures: leading-edge (LE) position of
EDCs selected at locally fitted $E_F$-MDC maxima (left), local extrema
of the fitted LE-position curves (middle), and local extrema of the
fitted EDC peak-position curves (right). Dashed horizontal lines
indicate the corresponding values at N. For comparison, the
rightmost panel shows the superconducting gap anisotropy previously
reported for termination B in Ref.~\cite{Changdar}. Despite the
different surface terminations and incomplete near-nodal coverage in
Ref.~\cite{Oleary}, the two independent datasets exhibit a similar
strongly anisotropic energy scale and comparable magnitude.
(c) Experimental quantities underlying the analysis in (b). Left:
raw EDCs extracted at the $E_F$-MDC maxima. Middle: fitted
leading-edge position versus momentum for the indicated cuts. Right:
fitted EDC peak position versus momentum. Numbers identify the
crossings marked in (a). Local fits are used only to determine the
positions of spectral features and do not assume a model for the full
ARPES spectral function.
}
    \label{fig:placeholder}
\end{figure*}

\begin{figure*}
    \centering
    \includegraphics[width=1\linewidth]{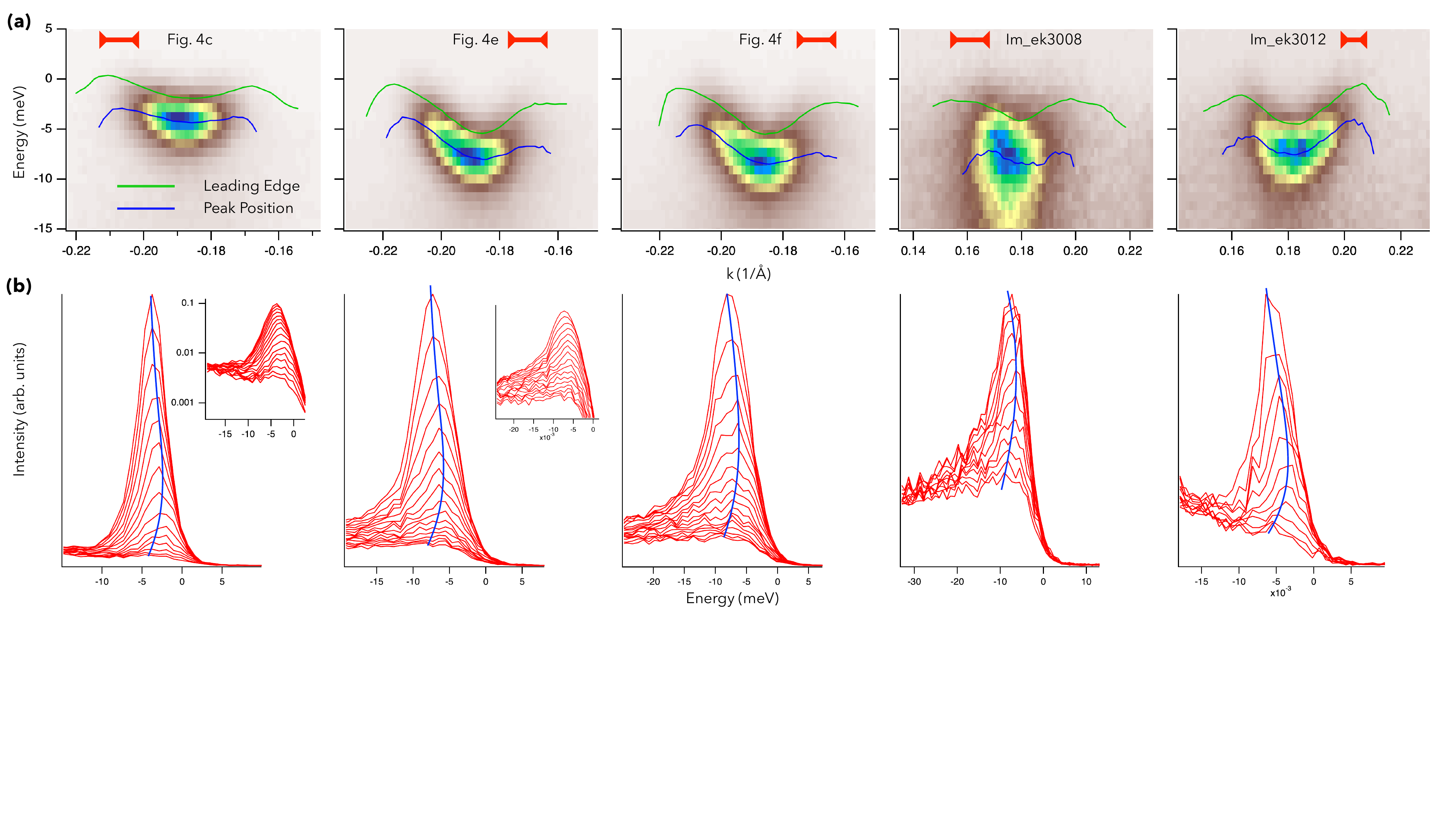}
    \caption{
\textbf{Back-bending of the dispersion in the deposited data of
O'Leary \textit{et al.}}
(a) Five representative momentum cuts from the deposited dataset.
The first three correspond to Figs.~4(c), 4(e), and 4(f) of
Ref.~\cite{Oleary}; Im\_ek3008 and Im\_ek3012 are additional
deposited cuts. Green and blue curves show, respectively, the
leading-edge and EDC peak positions obtained from local fits to each
EDC across the cuts. Both quantities reverse their dispersion near
the Fermi crossings, revealing back-bending on both sides of the
band. The red symbols at the top indicate the approximate momentum
intervals over which the back-bending occurs.
(b) Corresponding sequences of raw EDCs around one representative
Fermi crossing for each cut in (a). The spectra are shown without
Fermi-function division. Blue curves highlight the maxima of representative EDCs
near the reversal of the dispersion. Insets in the first two panels
show the same spectra on a logarithmic intensity scale. The
back-bending is therefore present directly in the evolution of the
experimental EDCs and is not a consequence of image contrast or
interpolation.
}
    \label{fig:placeholder}
\end{figure*}

\begin{figure*}
    \centering
    \includegraphics[width=1\linewidth]{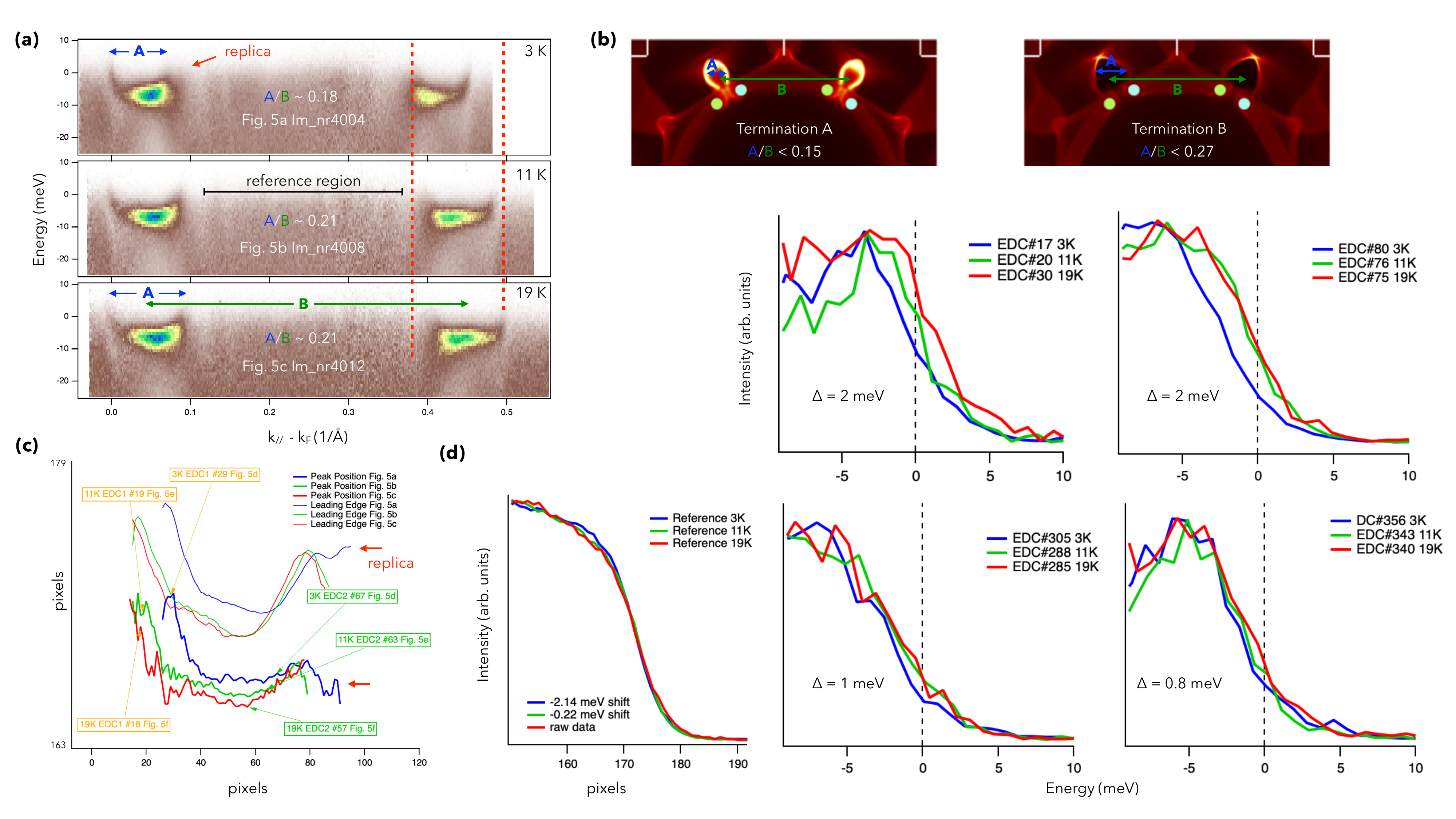}
    \caption{
\textbf{Reanalysis of the temperature-dependent data of
Ref.~\cite{Oleary}.}
(a) Complete deposited datasets underlying Figs.~5(a)--5(c) of
Ref.~\cite{Oleary}, at 3, 11, and 19 K, shown over their full
rescaled momentum ranges. $A$ denotes the characteristic arc width
and $B$ the separation of neighboring arcs. The resulting ratios are
$A/B\simeq0.18$, 0.21, and 0.21 at 3, 11, and 19 K, respectively.
The 3-K data additionally show a replica of the left arc.
Dashed vertical lines illustrate the relative displacement of
spectral structures between the three measurements. The region
between the arcs used for relative energy registration is indicated.
(b) Calculated Fermi-surface maps for terminations A and B from Ref.~\cite{Veyrat}, with the
same geometrical quantities $A$ and $B$ indicated. Termination A
satisfies $A/B<0.15$, whereas termination B allows substantially
larger values. The ratios measured in the Fig.~5 datasets are
incompatible with termination A and identify these measurements with
termination B.
(c) EDC peak-position and leading-edge-position curves extracted from
the 3-, 11-, and 19-K deposited arrays and plotted versus their
original angular and energy pixel coordinates. Labels identify the original
pixel numbers of the EDCs used in Fig.~5 of Ref.~\cite{Oleary} as
the Fermi-crossing and reference spectra. The curves demonstrate both
the relative displacement of the measurements and that the selected
reference EDCs do not represent equivalent positions of the three
dispersions.
(d) Relative energy registration and resulting temperature evolution. Lower left: Fermi cutoffs obtained from the metallic spectral weight in the reference region marked in (a), used to determine the relative energy shifts between temperatures. The absolute 19-K energy reference is retained from Ref.~\cite{Oleary}. Remaining panels: raw, unsmoothed EDCs at the four $E_F$-MDC maxima after relative registration. No individual EDCs are shifted independently. Original EDC pixel numbers are given in
the legends. Dashed vertical lines denote the surface-state Fermi
reference. The corresponding low-temperature gap values are
approximately $\Delta=2$, 2, 1, and 0.8 meV. All four crossings show
a systematic temperature evolution, while finite off-nodal
low-energy suppression remains visible already at 19 K.
}
    \label{fig:placeholder}
\end{figure*}

\setcounter{figure}{0}
\renewcommand{\thefigure}{S\arabic{figure}}

\begin{figure*}
    \centering
    \includegraphics[width=1\linewidth]{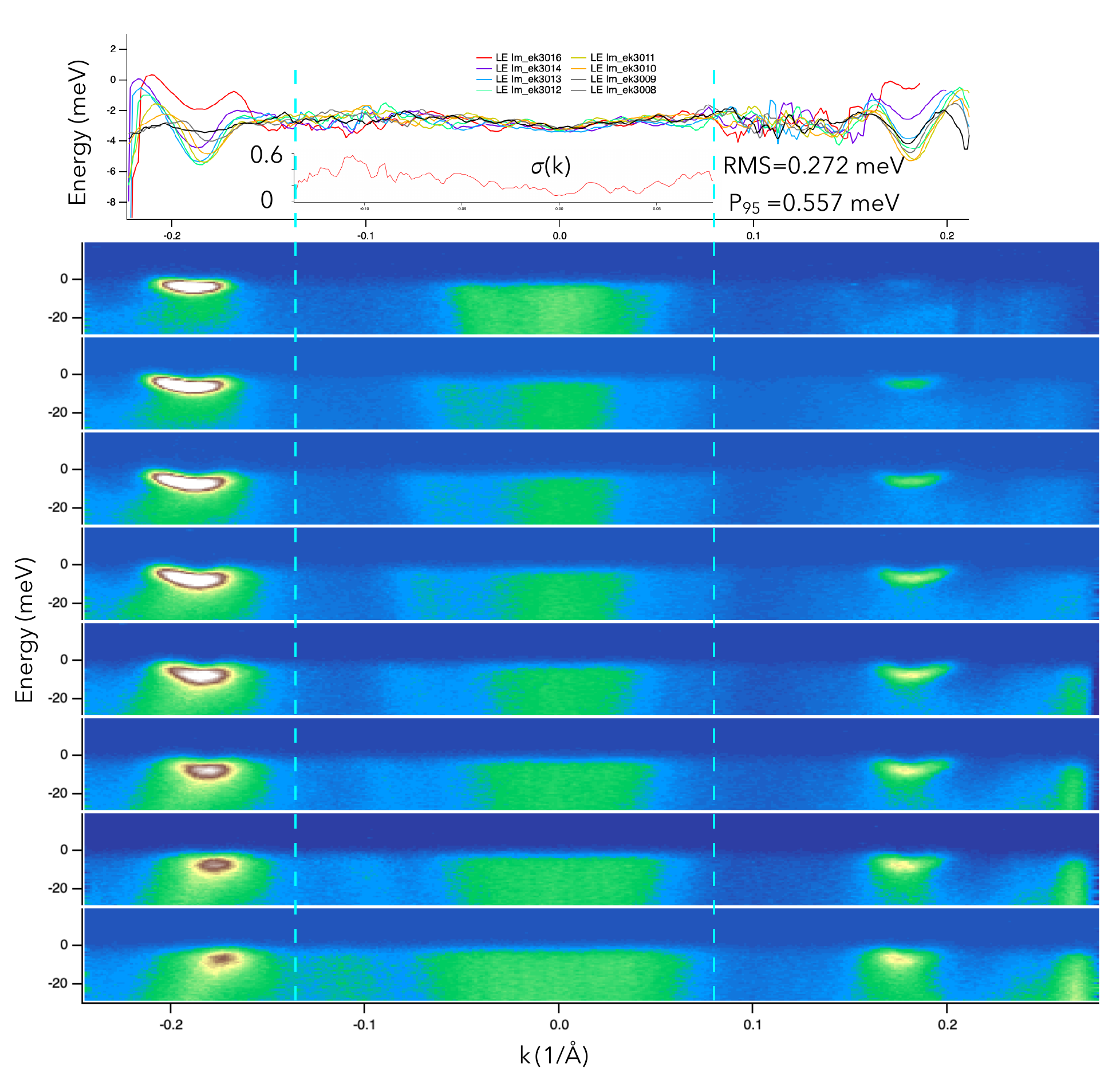}
    \caption{
\textbf{Internal test of relative energy reproducibility in the datasets underlying Fig.~4 of Ref.~\cite{Oleary}.}
Near-$E_F$ portions of the eight independently acquired deposited $I(E,k)$ datasets are shown together with leading-edge positions extracted from the bulk-derived metallic spectral weight. Leading edges are determined from local fits to the maxima of $dI/dE$. The upper panel compares the eight resulting trajectories; the inset shows their point-by-point standard deviation $\sigma(k)$. Detector channels 79--234 are used for the quantitative analysis, providing a broad interval of usable bulk-derived spectral weight while excluding the Fermi-arc features. Within this interval, the rms deviation of the individual leading-edge positions from their point-by-point mean is 0.272~meV. Of the 1248 residuals, 95\% satisfy $|\delta E|<0.557$~meV. The mean offsets of the eight independently acquired cuts from their common trajectory range from $-0.154$ to $+0.132$~meV.
}
    \label{fig:S1}
\end{figure*}
\end{document}